\documentclass{elsart}
\usepackage{amsmath}
\usepackage{amssymb}
\usepackage{graphicx}
\journal{Physica A}

\providecommand{\msii}{\mbox{m/s$^2$}\ }
\providecommand{\refkl}[1]{(\ref{#1})}

\providecommand{\sub}[1]{_{\rm #1}}
\renewcommand{\sup}[1]{^{\rm #1}}
\providecommand{\erw}[1]{\mbox{$\langle #1 \rangle$}}

\begin{document}
\begin{frontmatter}

\title{Delays, Inaccuracies and Anticipation in Microscopic Traffic Models}
\author{Martin Treiber},  \ead{treiber@vwi.tu-dresden.de} \ead[url]{http://www.mtreiber.de}
\author{Arne Kesting}, \ead{kesting@vwi.tu-dresden.de} \ead[url]{http://www.akesting.de}
\author{Dirk Helbing} \ead{helbing1@vwi.tu-dresden.de}\ead[url]{http://www.helbing.org}

\address{Institute for Transport \& Economics, 
  Dresden University of Technology, 
  Andreas-Schubert-Strasse 23, D-01062 Dresden, Germany}

\date{2 May 2005}
\begin{abstract}
We generalize a wide class of time-continuous microscopic traffic
models to include essential aspects of driver behaviour not captured
by these models.  Specifically, we consider (i) finite reaction times,
(ii) estimation errors, (iii) looking several vehicles ahead (spatial
anticipation), and (iv) temporal anticipation.  The estimation errors
are modelled as stochastic Wiener processes and lead to
time-correlated fluctuations of the acceleration.

We show that the destabilizing effects of reaction times and
estimation errors can essentially be compensated for by spatial and
temporal anticipation, that is, the combination of stabilizing and
destabilizing effects results in the same qualitative macroscopic
dynamics as that of the respectively underlying simple car-following
model. In many cases, this justifies the use of simplified,
physics-oriented models with a few parameters only. Although the
qualitative dynamics is unchanged, multi-anticipation increase both
spatial and temporal scales of stop-and-go waves and other complex
patterns of congested traffic in agreement with real traffic data.
Remarkably, the anticipation allows accident-free smooth driving in
complex traffic situations even if reaction times exceed typical time
headways.
\end{abstract}
\end{frontmatter}

\section{\label{sec:intro}Introduction}
The nature of human driving behaviour and the differences with the automated
driving implemented in most micromodels is a controversial topic in
traffic science \cite{Helb-Opus,Brackstone-hist,Holland97,Helb-crit,nagel-wagner-vdiff,Kerner-traffic,chowdhury-report,nagatani-report}.
Finite reaction times and estimation capabilities
impair the human driving performance and stability
compared to automated driving, sometimes called `adaptive cruise
control' (ACC).  However, unlike machines, human drivers routinely
scan the traffic situation several vehicles ahead and anticipate
future traffic situations leading, in turn, to an increased stability.

The question arises, how this behaviour affects the overall driving
behaviour and performance, and whether the stabilizing effects (such
as anticipation) or the destabilizing effects (like reaction times and
estimation errors) dominate, or if they effectively cancel out each
other.  The answers to these questions are crucial for determining the
influence of a growing number of vehicles equipped with automated
acceleration control on the overall traffic flow. Up to now, there is
not even clarity about the sign of the effect.  Some investigations
predict a positive effect \cite{Treiber-aut}, while others are more
pessimistic \cite{marsden-ACC}.

Single aspects of human driving behaviour have been investigated in the
past.  For example, it is well known that traffic instabilities
increase with the reaction times $T'$ of the drivers. 
Finite reaction times in time-continuous models are implemented by
evaluating the right-hand side of the equation for the acceleration
(or velocity) at some previous time $t-T'$ with $T'>0$
\cite{Newell,Bando-98,Davis-02}. Reaction times in
time-continuous models have been modelled as early as 1961 by Newell
\cite{Newell}. Recently, the optimal-velocity model (OVM)
\cite{Bando} has been extended to include finite reaction times
\cite{Bando-98}. 
However, the Newell model has no dynamic velocity, and the OVM with
delay turns out to be accident-free only for unrealistically small
reaction times \cite{BandoDelay-02}.

To overcome this deficiency, Davis \cite{Davis-02} has introduced
(among other modifications) an anticipation of the expected future gap
to the front vehicle allowing accident-free driving at reaction times
of 1 s.  However, reaction times were not fully implemented in
Ref.~\cite{Davis-02} since the own velocity, which is one of the
stimuli on the right-hand side of the acceleration equation, has been
taken at the actual rather than at the delayed time.

Another approach to model temporal anticipation consists in including
the acceleration of the preceding vehicles in the input variables of
the model. For cellular automata, this has been implemented by
introducing a binary-valued `brake light' variable \cite{Kno01}.

To our knowledge, there exists no car-following model exhibiting
platoon stability (with respect to all stimuli) for reaction times
exceeding half of the time headway of the platoon vehicles.  Human
drivers, however, accomplish this task easily: In dense (not yet
congested) traffic, the most probable time headways on German freeways
are 0.9-1s \cite{Tilch-TGF99,Kno02-data} which is of the same order as
typical reaction times \cite{green-reactionTimes}.  However,
single-vehicle data for German freeways
\cite{Tilch-TGF99,Kno02-data} indicate that some drivers drive at
headways as low as 0.3 s, which is below the reaction time of even a
very attentive driver by a factor of at least 2-3
\cite{green-reactionTimes}. For principal reasons, therefore, safe
driving is not possible in this case when considering only the
immediate vehicle in front.

This suggests that human drivers achieve additional stability and
safety by taking into account next-nearest neighbors and further
vehicles ahead as well. Such `spatial anticipation' or
`multi-anticipation' has been applied to the OVM \cite{Lenz-Wagner}
and to the Gipps model \cite{Eissfeldt03} as well as to some cellular
automata model \cite{Kno01,CA_limitedAcc}.  As expected, the resulting
models show a higher stability than the original model.  However, the
stability of the aforementioned models is still smaller than that of
human driving. Furthermore, they display unrealistic behaviour such as
clustering in pairs \cite{Eissfeldt03}, or too elevated propagation
velocities of perturbations in congested traffic ($v_g=-30$ km/h)
\cite{Lenz-Wagner}.

Imperfect estimation capabilities often serve as motivation or
justification to introduce stochastic terms into micromodels such as
the Gipps model \cite{Gipps81} (see also \cite{Brackstone-hist}).
Most cellular automata require fluctuating terms as well.  In nearly
all the cases, fluctuations are assumed to be $\delta$-correlated in
time and acting directly on the accelerations.  An important feature
of human estimation errors, however, is a certain {\it persistency}.
If one underestimates, say, the distance at time $t$, the probability
of underestimating it at the next time step (which typically is less
than 1 s in the future) is high as well.  Another source leading to
temporally correlated acceleration noise lies in the concept of
`action points' modelling the tendency of human drivers to actively
adapt to the traffic situation, i.e., to change the acceleration only
at discrete times \cite{Wagner-03}.

In this paper, we propose the human driver (meta-)model (HDM) in terms
of four extensions to basic physics-oriented traffic models
incorporating into these models (i) finite reaction times, (ii)
estimation errors, (iii) spatial anticipation, and (iv) temporal
anticipation. The class of suitable basic models is characterized by
continuous acceleration functions depending on the velocity, the gap,
and the relative velocity with respect to the preceding car and
includes, for example, the optimal-velocity model (OVM)
\cite{Bando}, the Gipps model \cite{Gipps81}, the velocity-difference
model \cite{Jiang-vdiff}, the intelligent-driver model (IDM)
\cite{Opus}, and the boundedly rational driver model
\cite{Lub03,Mahnke03}.

For matters of illustration, we will apply the HDM to the
intelligent-driver model \cite{Opus}, which has a built-in
anticipative and smooth braking strategy, and which reaches good
scores in a first independent attempt to benchmark micromodels based
on real traffic data \cite{Brockfeld-Benchmark}.

In Sec. \ref{sec:model}, we will formulate the HDM in terms of the
acceleration function of the basic model. In Sec. \ref{sec:results} we
will simulate the stability of vehicle platoons as a function of the
reaction time $T'$ and the number of anticipated vehicles $n_a$. We
find string stability for arbitrarily long platoons for reaction times
of up to 1.5 s.  Furthermore, we simulate the macroscopic traffic
dynamics for an open system containing a flow-conserving bottleneck
\cite{Opus,Treiber-TGF99}. We find that multi-vehicle anticipation
($n_a>1$) can compensate for the destabilizing effects of reaction
times and estimation errors.  The numerically determined phase diagram
for the corresponding parameter space gives the conditions under which
simple physical models describe the traffic dynamics correctly. In the
concluding Section \ref{sec:diss}, we suggest applications and further
investigations and discuss some aspects of human driving that are not
included in the HDM.

\section{\label{sec:model}Modelling human driver behaviour}
%
Let us formulate the HDM as a meta-model applicable to 
time-continuous
micromodels (car-following models) of the general form
\begin{equation}
\label{mic}
\frac{dv_{\alpha}}{dt}  = a\sup{mic}\left(
s_{\alpha}, v_{\alpha}, 
\Delta v_{\alpha}\right),
\end{equation}
where the own velocity $v_{\alpha}$, the net distance $s_{\alpha}$,
and the velocity difference $\Delta v_{\alpha}$ to the leading vehicle
serve as stimuli determining the acceleration $a\sup{mic}$
\cite{remark-acc}.  This class of basic models is characterized by (i)
instantaneous reaction, (ii) reaction only to the immediate
predecessor, and (iii) infinitely exact estimating capabilities of
drivers regarding the input stimuli $s$, $v$, and $\Delta v$, which
also means that there are no fluctuations. In some sense, such models
describe driving behaviour similar to adaptive cruise control systems.

For the sake of simplicity, we will restrict ourselves to single-lane
longitudinal dynamics. Furthermore, we will not include adaptations of
drivers to the traffic conditions of the last few minutes. This
so-called `memory effect' is discussed elsewhere
\cite{IDMM}.

\subsection{\label{sec:react}Finite reaction time}
A reaction time $T'$ is implemented simply by evaluating the
right-hand side of Eq. \refkl{mic} at time $t-T'$. If $T'$ is not a
multiple of the update time interval, we propose a linear
interpolation according to
\begin{equation}
\label{intp}
x(t-T')=\beta x_{t-n-1}+(1-\beta) x_{t-n},
\end{equation}
where $x$ denotes any quantity on the right-hand side of \refkl{mic}
such as $s_{\alpha}$, $v_{\alpha}$, or $\Delta v_{\alpha}$, and
$x_{t-n}$ denotes this quantity taken $n$ time steps before the actual
step. Here, $n$ is the integer part of $T'/\Delta t$, and the weight
factor of the linear interpolation is given by $\beta=T'/\Delta t-n$.
We emphasize that \textit{all} input stimuli $s_{\alpha}$,
$v_{\alpha}$, and $\Delta v_{\alpha}$ are evaluated at the delayed
time.

Notice that the reaction time $T'$ is sometimes set equal to the
`safety' time-headway $T$.  It is, however, essential to distinguish
between these times conceptually. While the time headway $T$ is a
characteristic parameter of the driving style, the reaction time $T'$
is essentially a physiological parameter and, consequently, at most
weakly correlated with $T$.  We point out that both the time headway
$T$ and the reaction time $T'$ are to be distinguished from the
numerical update time step $\Delta t$, which is sometimes erroneously
interpreted as a reaction time as well.  For example, in our
simulations, an update time step of 2 s has about the same effect as a
reaction time of 1 s while the results are essentially identical for
any update time step below 0.2 s.

\subsection{\label{sec:fluct}Imperfect estimation capabilities}
We will now model estimation errors for the net distance $s$ and the
velocity difference $\Delta v$ to the preceding vehicle. Since the
velocity itself can be obtained by looking at the speedometer, we
neglect its estimation error.  From empirical investigations (for an
overview see \cite{Brackstone-hist}, p. 190) it is known that the
uncertainty of the estimation of $\Delta v$ is proportional to the
distance, i.e., one can estimate the time-to collision
(TTC) $s/|\Delta v|$ with a constant uncertainty \cite{remark-bias}.
For the distance itself, we specify the estimation error in a relative
way by assuming a constant variation coefficient $V_s$ of the
errors.  Furthermore, in contrast to other stochastic micromodels 
\cite{Nagel-S}, we take into account a finite persistence of
estimation errors by modelling them as a Wiener process
\cite{Gardiner}.  This leads to the following nonlinear stochastic
processes for the distance and the velocity difference,
\begin{equation}
s\sup{est}(t)=s(t)\,\exp (V_s w_s(t)), 
\end{equation}
\begin{equation}
(\Delta v)\sup{est}(t)= \Delta v(t) + s(t)\,r\sub{c} w_{\Delta v}(t),
\end{equation}

where $V_s=\sigma_s/\erw{s}$ with $\sigma_s^2=\erw{(s-\erw{s})^2}$ is
the variation coefficient of the distance estimate, and $1/r\sub{c}$,
a measure for the average estimation error of the time to collision.
The stochastic variables $w_s(t)$ and $w_{\Delta v}(t)$ obey
independent Wiener processes $w(t)$ of variance 1 with correlation
times $\tau=1$ defined by \cite{Gardiner}
\begin{equation}
\label{Wiener}
\frac{dw}{dt}=- \frac{w}{\tau} + \sqrt{\frac{2}{\tau}}\,\xi(t),
\end{equation}
with
\begin{equation}
\label{xi}
\erw{\xi}=0, \ \ \erw{\xi(t)\xi(t')}= \delta(t-t').
\end{equation}

In the explicit numerical update from time step $t$ to step $t+\Delta t$, 
we implemented the Wiener processes by the approximations
\begin{equation}
\label{WienerNum}
w_{t+\Delta t} = e^{-\Delta t/\tau} w_t + \sqrt{\frac{2 \Delta t}{\tau}} \eta_t,
\end{equation}
where the $\{\eta_t\}$ are independent realizations of a Gaussian
distributed quantity with zero mean and unit variance.  We have
checked numerically that the update scheme \refkl{WienerNum} satisfies
the fluctuation-dissipation theorem $\erw{w_t^2}=1$ for any update
time interval satisfying $\Delta t\ll \tau$.

Simulations have shown that, in agreement with expectation, traffic
becomes more unstable with increasing values of $V_s$ and $r_c$.  To
compare the influence of the temporally correlated multiplicative HDM
noise with more conventional white acceleration noise, we have
repeated the simulations of Section \ref{sec:results} with the
deterministic HDM ($V_s=r_c=0$) augmented by an additive noise term
$\sqrt{Q_a}\xi(t)$ at the right-hand side of the acceleration
equation.  Remarkably, the dynamics did not change essentially for
reasonable values of the fluctuation strength $Q_a$. Thus, the more
detailed representation of stochasticity by the HDM can be used to
relate the conventional noise strength $Q$ (which does not have any
intuitive meaning) to better justified noise sources.

\subsection{Temporal anticipation}
We will assume that drivers are aware of their finite reaction time
and anticipate the traffic situation accordingly.  Besides
anticipating the future distance \cite{Davis-02}, we will anticipate
the future velocity using a {\it constant-acceleration heuristics}.  The
combined effects of a finite reaction time, estimation errors and
temporal anticipation leads to the following input variables for the
underlying micromodel \refkl{mic}:
\begin{equation}
\frac{dv}{dt}=a\sup{mic}(s'_{\alpha}, v'_{\alpha}, \Delta v'_{\alpha})
\end{equation}
with
\begin{equation}
\label{s_est}
s'_{\alpha}(t)=\left[s_{\alpha}\sup{est} 
  -T' \Delta v_{\alpha}\sup{est}\right]_{t-T'},
\end{equation}
\begin{equation}
\label{v_est}
v'_{\alpha}(t)=\left[v_{\alpha}\sup{est} + T' a_{\alpha}\right]_{t-T'},
\end{equation}
and
\begin{equation}
\label{dv_est}
\Delta v'_{\alpha}(t)=\Delta v_{\alpha}\sup{est}(t-T').
\end{equation}
We did not apply the constant-acceleration heuristics for the
anticipation of the future velocity difference or the future distance,
as the accelerations of other vehicles cannot be estimated reliably
by human drivers.  Instead, we have applied the simpler {\it
constant-velocity heuristics} for these cases.

Notice that the anticipation terms discussed in this subsection (which
do not contain any additional model parameters) are specifically
designed to compensate for the reaction time by means of plausible
heuristics. They are to be distinguished from `anticipation' terms in
some models aiming at collision-free driving in `worst-case' scenarios
(sudden braking of the preceding vehicle to a standstill) when the
braking deceleration is limited.  Such terms typically depend on the
velocity difference and are included, e.g., in the Gipps model, in the
IDM, and in some cellular automata \cite{CA_limitedAcc,Knospe}, but
notably not in the OVM.  The HDM is most effective when using a basic
model with this kind of anticipation.

\subsection{Spatial anticipation for several vehicles ahead}
Let us now split up the acceleration of the underlying microscopic
model into a single-vehicle acceleration on a nearly empty road
depending on the considered vehicle $\alpha$ only, and a braking
deceleration taking into account the vehicle-vehicle interaction with
the preceding vehicle:
\begin{equation}
\label{a_free_int}
a\sup{mic}(s_{\alpha}, v_{\alpha}, \Delta v_{\alpha})
:= a\sup{free}_{\alpha}
  + a\sup{int}(s_{\alpha}, v_{\alpha}, \Delta v_{\alpha}).
\end{equation}
Notice that this decomposition of the acceleration has already been
used to formulate a lane-changing model for a wide class of
micromodels \cite{MOBIL-rostock}.

Next, we model the reaction to several vehicles ahead just by summing
up the corresponding vehicle-vehicle pair interactions
$a\sup{int}_{\alpha\beta}$ from vehicle $\beta$ to vehicle $\alpha$
for the $n\sub{a}$ nearest preceding vehicles $\beta$:
\begin{equation}
\label{a_nonloc}
\frac{d}{dt} v_{\alpha}(t) = a\sup{free}_{\alpha} 
  + \sum_{\beta=\alpha-n\sub{a}}^{\alpha-1}
   a\sup{int}_{\alpha \beta},
\end{equation}
where all distances, velocities and velocity differences on the
right-hand side are given by \refkl{s_est} - \refkl{dv_est}.  Each
pair interaction between vehicle $\alpha$ and vehicle $\beta$ is
specified by
\begin{equation}
a\sup{int}_{\alpha\beta}
=a\sup{int} \left(s_{\alpha \beta}, v_{\alpha}, v_{\alpha}-v_{\beta}\right),
\end{equation}
where
\begin{equation}
s_{\alpha \beta}=
  \sum_{j=\beta+1}^{\alpha} \!\! s_j
\end{equation}
is the sum of all {\it net} gaps between the vehicles $\alpha$ 
and $\beta$.

\subsection{\label{sec:IDM}Applying the HDM extensions to the intelligent driver model (IDM)}
In this paper, we will apply the HDM extensions to the IDM.  In
this model \cite{Opus}, the acceleration of each vehicle $\alpha$ is
assumed to be a continuous function of the velocity $v_{\alpha}$, the
net distance gap $s_{\alpha}$, and the velocity difference
(approaching rate) $\Delta v_{\alpha}$ to the leading vehicle:
\begin{equation}
\label{IDMv}
\dot{v}_{\alpha} = a
         \left[ 1 -\left( \frac{v_{\alpha}}{v_0} 
                  \right)^4 
                  -\left( \frac{s^*(v_{\alpha},\Delta v_{\alpha})}
                                {s_{\alpha}} \right)^2
         \right].
\end{equation}
The IDM acceleration consists of a free acceleration 
 $a\sup{free} = a[1-(v/v_0)^4]$ for approaching  the desired velocity $v_0$
with an acceleration slightly below $a$, and the braking interaction
$a\sup{int} = -a(s^*/s)^2$, where the actual gap $s_{\alpha}$ is
compared with the `desired
minimum gap'
\begin{equation}
\label{sstar}
s^*(v, \Delta v) 
    = s_0 
    + v T
    + \frac{v \Delta v }  {2\sqrt{a b}},
\end{equation}
which is specified by the sum of the minimum distance $s_0$, the
velocity-dependent safety distance $v T$ corresponding to the time
headway $T$, and a dynamic part. The dynamic part implements an
accident-free `intelligent' braking strategy that, in nearly all
situations, limits braking decelerations to the `comfortable
deceleration' $b$.  Notice that all IDM parameters have an intuitive
meaning. By an appropriate scaling of space and time, the number of
parameters can be reduced from five to three.

\subsubsection*{Renormalisation for the intelligent driver model (IDM)}
Remarkably, there exists a closed-form solution of the
multi-anticipative IDM equilibrium distance as a function of the velocity,
\begin{equation}
\label{seq}
s_{e}(v) = 
\gamma s^*(v,0) 
  \left[ 1 - \left(\frac{v}{v_0}\right)^{\delta}\right]^{-\frac{1}{2}},
\end{equation}
which is $\gamma$ times the equilibrium distance
of the original IDM \cite{Opus}, where
\begin{equation}
\label{sumii}
\gamma=\sqrt{\sum_{\alpha=1}^{n\sub{a}} \frac{1}{\alpha^2}}.
\end{equation}
The equilibrium distance $s_e(v)$ can be transformed
to that of the original IDM by renormalizing the relevant IDM parameters
appearing in $s^*(v,0)$:
\begin{equation}
s\sup{ren}_0=\frac{s_0}{\gamma}, \ \ T\sup{ren}=\frac{T}{\gamma}.
\end{equation}
The above renormalisation will be applied to all simulations of this
paper. Notice that, in the limiting case of anticipation to
arbitrarily many vehicles we obtain $\lim_{n\sub{a}\to\infty}
\gamma(n\sub{a})=\pi/\sqrt{6} = 1.283$.  This means that the combined
effects of all non-nearest-neighbor interactions would lead to an
increase in the equilibrium distance by just about 28\%.

\begin{table}
\begin{center}
\begin{tabular}{ll} 
 Parameter                                           & Value \\[1mm] \hline 
\\[-2mm]
 Reaction time $T'$                                  & 0 s - 2.0 s \\
 Number of anticipated vehicles $n_a$                & 1 -- 7 \\
 Relative distance error $V_s$                       & 5\% \\
 Inverse TTC error $r_c$                             & 0.01/s \\
 Error correlation time $\tau$                       & 20 s 
\end{tabular}
\caption{\label{tab:HDM}Parameters of the human-driver extensions 
with the values used in this paper. Unless stated otherwise, we have used
the IDM parameters $v_0=128$ km/h, $T=1.1$ s, $a=1 \ \msii$, $b=1.5 \
\msii$, and $s_0=2$ m. In Section \protect\ref{sec:platoon}, we have changed $v_0$ and $T$ to
115 km/h and 1.5 s, respectively.}
\end{center}
\end{table}

\subsection{Summary of the human driver model (HDM)}
The HDM is formulated in terms of a meta-model introducing reaction
times, finite estimation capabilities, temporal anticipation and
multi-vehicle anticipation to a wide class of simple micromodels.  The
model has two deterministic parameters, namely the reaction time $T'$
and the number $n_a$ of anticipated vehicles whose influences will be
investigated below.

The only stochastic contributions come from modeling finite estimation
capabilities.  The stochastic sources $V_s$ and $r_c$ characterize the
degree of the estimation uncertainty of the drivers, while $\tau$
denotes the correlation time of errors.  The limit $\tau \to 0$
corresponds to multiplicative white acceleration noise while $\tau \to
\infty$ corresponds to `frozen' error amplitudes, i.e., {\it de facto}
heterogeneous traffic. All human-driver extensions are switched off
and the original basic model is recovered if $T'=0$, $n_a=1$, and
$V_s=r_c=0$.

The HDM-IDM combination (i.e. the application of the HDM to the IDM)
has a total number of ten parameters which can be reduced to eight by
an appropriate scaling of space and time. Replacing the HDM noise by
white additive noise (cf. Sec. \ref{sec:fluct}) allows a further
reduction to six parameters while retaining all essential
properties. 

\section{\label{sec:results}Simulations and results}
In this section, we apply the HDM extensions to the IDM for matters of
illustration. In all simulations, we have used an explicit integration
scheme assuming constant accelerations between each update time
interval $\Delta t$ according to
\begin{equation}
\label{euler2}
\begin{array}{l}
v_{\alpha}(t+\Delta t)=v_{\alpha}(t)+ a_{\alpha}(t) \Delta t,\\
x_{\alpha}(t+\Delta t)=x_{\alpha}(t)+ v_{\alpha}(t) \Delta t
+ \frac{1}{2} a_{\alpha}(t) (\Delta t)^2.
\end{array}
\end{equation}
Unless stated otherwise, we will use the IDM and HDM parameters given
in Table \ref{tab:HDM}.  In the simulations, we will mainly study the
influences of the reaction time $T'$ and the number $n_a$ of
anticipated vehicles.

\subsection{\label{sec:platoon}String stability of a platoon}
We have investigated the stability of the HDM as a function of the
reaction time $T'$ and the number $n_a$ of anticipated vehicles by
simulating a platoon of $100$ vehicles following an externally
controlled lead vehicle.  As in a similar study for the OVM
\cite{BandoDelay-02,Davis-02}, the lead vehicle drives at
$v\sub{lead}=15.34\:\mathrm{m/s}$ for the first 1000 s before it
decelerates with $-0.7\:\mathrm{m/s}^2$ to $14.0\:\mathrm{m/s}$ and
continues with this velocity until the simulation ends at
$2500\:\mathrm{s}$.

For the platoon vehicles, we use the IDM parameters
$v_0=32\:\mathrm{m/s}$ and $T=1.5\:\mathrm{s}$ to obtain the same
desired velocity and initial equilibrium gap
($s\sub{e}=25.7\:\mathrm{m}$) as in previous studies
\cite{BandoDelay-02,Davis-02}.  The other IDM parameters are 
$a=1\,\msii$, $b=1.5\,\msii$, and $s_0=2$ m.  If $n_a$ is larger than
the number of preceding vehicles (which can happen for the first
vehicles of the platoon) then $n_a$ is reduced accordingly.
Fluctuations have been neglected in this scenario. As initial
conditions, we have assumed the platoon to be in equilibrium, i.e.,
the initial velocities of all platoon vehicles were equal to
$v\sub{lead}$ and the gaps equal to $s_e$ so that the initial HDM (and
IDM) accelerations were equal to zero.

We distinguish three stability regimes: (i) String stability, i.e.,
all perturbations introduced by the deceleration of the lead vehicles
are damped away, (ii) an oscillatory regime, where perturbations
increase but do not lead to crashes, and (iii) an instability with
accidents.  The condition for a simulation to be in the crash regime
(iii) is fulfilled if there is {\it some} time $t$ and {\it some}
vehicle $\alpha$ so that $s_{\alpha}(t)<0$. The condition for string
stability is fulfilled if $|\dot{v}_{\alpha}(t)|<2\, \msii$ at {\it
all} times (including the period where the leading vehicle
decelerates) and for {\it all} vehicles, and additionally
$|\dot{v}_{\alpha}(t)|<0.01\,
\msii$ for all vehicles towards the end of the simulation. Finally, if
neither the conditions for the crash regime nor that for the stable
regime are fulfilled, the simulation result is attributed to the
oscillatory regime.

\begin{figure}
  \begin{center}
    \includegraphics[width=60mm]{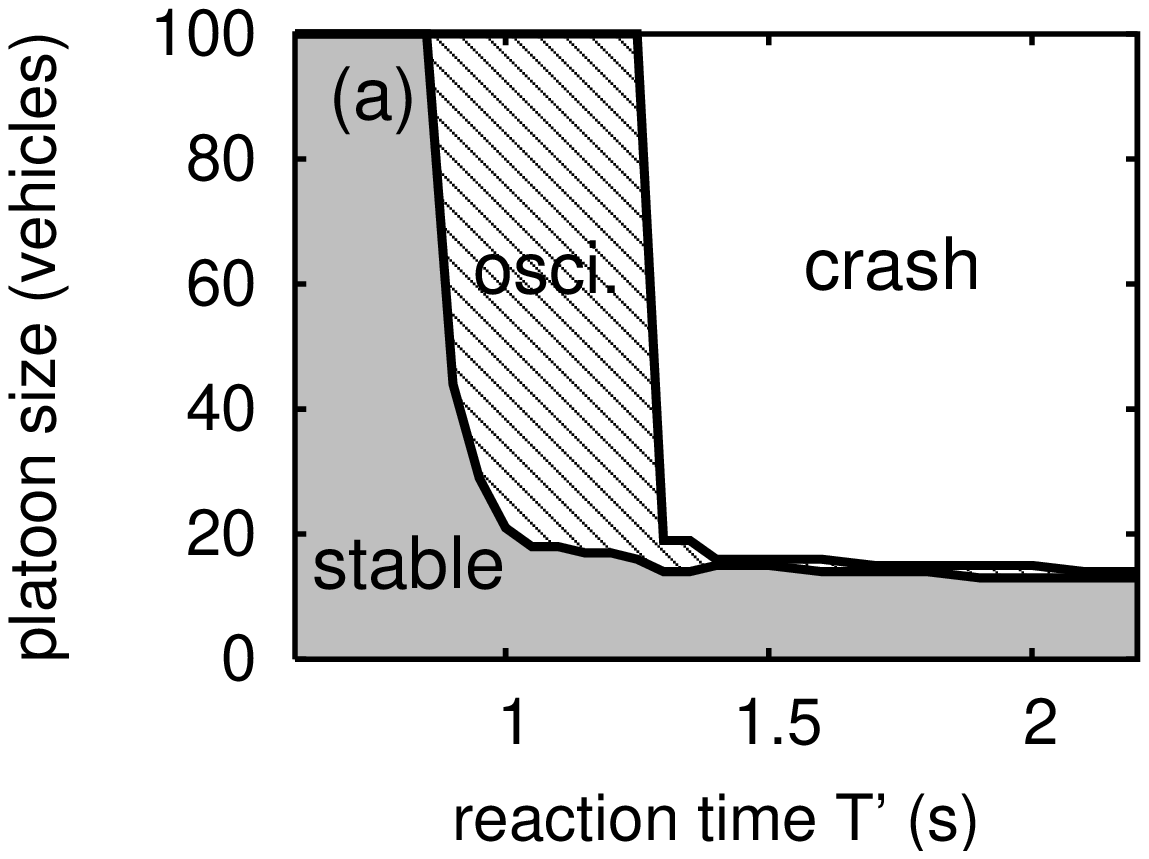}
    \includegraphics[width=60mm]{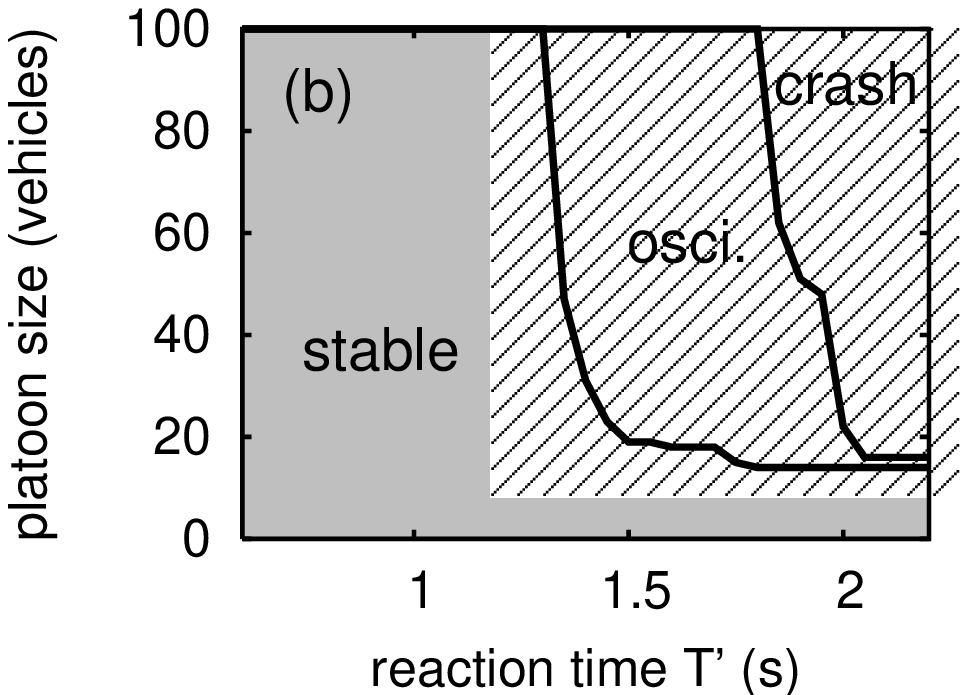}
  \end{center}

\caption{\label{figStringPlatoon} Stability of a platoon of identical vehicles
  as a function of the platoon size and the reaction time $T'$ for the
  situation described in Section \protect\ref{sec:platoon} (a) assuming conventional
  follow-the-leader behaviour ($n_a=1$) and (b) a reaction to
  $n_a=5$ vehicles. The simulation is for a time headway of
  $T=1.5\,\mathrm{s}$, a numerical update time interval of $\Delta
  t=0.1$ s, and a desired velocity $v_0=32\,\mathrm{m/s}$. The other
  parameters are given in Table \protect\ref{tab:HDM}. In
  the `stable' phase, all perturbations are damped away. In the
  oscillatory regime, the perturbations increase, but do not lead to
  crashes.  }
\end{figure}

Figure \ref{figStringPlatoon} shows the three stability regimes as a
function of the reaction time $T'$ and the platoon size $n$ for
spatial anticipations of $n_a=1$ and $5$ vehicles, respectively. For
$n_a=1$ (corresponding to conventional car-following models without
spatial anticipation), a platoon of $100$ vehicles is stable for
reaction times of up to $T'\sub{c1}=0.8\:\mathrm{s}$. Test runs with
larger platoon sizes (up to 1000 vehicles) did not result in different
thresholds suggesting that stability for a platoon size of 100
essentially means stability for arbitrarily large platoon sizes.

Increasing the spatial anticipation to $n_a=5$ vehicles shifted the
threshold of the delay time $T'$ for string stability of a platoon of
100 vehicles to $T'\sub{c1}=1.3\:\mathrm{s}$.  Increasing the delay
time $T'$ beyond the stability threshold led to strong oscillations
of the platoon. Crashes, however, occurred only when $T'$ exceeded a
second threshold $T'\sub{c2}$.  Remarkably, for $n_a=5$ or more
vehicles, the observed threshold $T'\sub{c2}=1.8\:\mathrm{s}$ is
larger than the equilibrium time headway $s_e/v\sub{lead}=1.68$ s.
More detailed investigations reveal that crashes are triggered
either directly by late reactions to deceleration maneuvers or
indirectly as a consequence of the string instability.  Further
increasing $n_a$ do not change the thresholds significantly.

%

\subsection{\label{simulation-B}Open system with a bottleneck}
In this section we examine the opposite influences of the driver
reaction time $T'$ and the spatial anticipation $n_a$ on the stability of
traffic and the occurring traffic states in a more complex and
realistic situation.

We have simulated a single-lane road section of total length 20 km
with a bottleneck and open boundaries assuming identical drivers and
vehicles of length $l=5\:\mathrm{m}$, whose parameters are given in
Table \ref{tab:HDM}.  The update time interval of the numerical
integration was $\Delta t=0.1\:\mathrm{s}$. Each simulation run
covered a time interval of $3\:\mathrm{h}$.  We initialized the
simulations with very light traffic of density 1 vehicle/km and
set all initial velocities to 100 km/h.

We have simulated idealized rush-hour conditions by increasing the
inflow $Q\sub{in}(t)$ at the upstream boundary linearly from 100 veh/h
at $t=0$ to 2100~veh/h at $t=1$~h, keeping the traffic demand constant
afterwards.  Since this demand exceeds the static road capacity $Q_B
\approx 2000$ veh/h at the bottleneck (the maximum of the fundamental
diagram), a traffic breakdown is always provoked, irrespective of the
stability of traffic.  We have implemented a flow-conserving
bottleneck at 18 km $\le x\le$ 20 km by linearly increasing the IDM
parameter $T$ from 1.1~s to 1.65~s in the region 18.0 km $\le x\le$
18.5 km, setting $T=1.65$ s for 18.5 km $\le x\le$ 19.5 km, and
linearly decreasing $T$ from 1.65 s to 1.1 s in the region 19.5 km
$\le x\le$ 20.0 km (see Ref. \cite{Helb-crit} for a justification of
this treatment of flow-conserving bottlenecks).

In order to determine the spatiotemporal dynamics, we plot, at any
given spatiotemporal point $(x,t)$, the locally averaged velocity of
the vehicle trajectories nearby. The averaging filter
\cite{Treiber-smooth} had half-widths of 1 minute and 0.4 kilometers,
respectively.

\begin{figure}
\begin{center}
  \includegraphics[width=100mm]{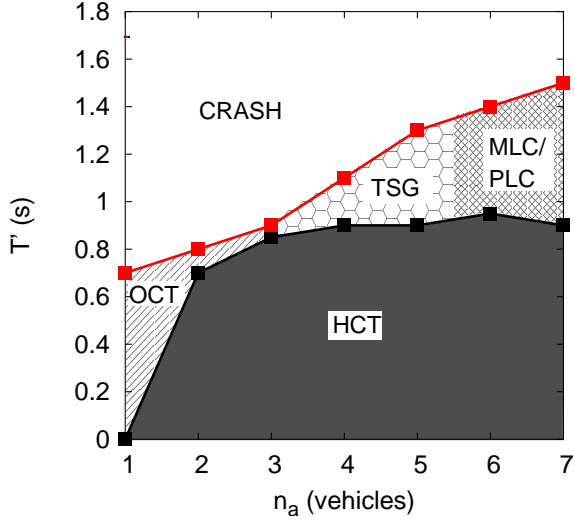}
\end{center}
\caption{\label{figPhaseDiagramTrafficStates} 
Phase diagram of congested traffic states in the phase space spanned by the
number $n_a$ of anticipated vehicles and the reaction time $T'$ in the
open system with a bottleneck as described in the text.  The dynamic
phases HCT (homogeneous congested traffic), OCT (oscillatory congested
traffic), TSG (triggered stop-and go), and MLC/PLC (moving and pinned
localized clusters) are discussed in the main text.}
\end{figure}

\begin{figure}
  \begin{center}
 \includegraphics[width=65mm]{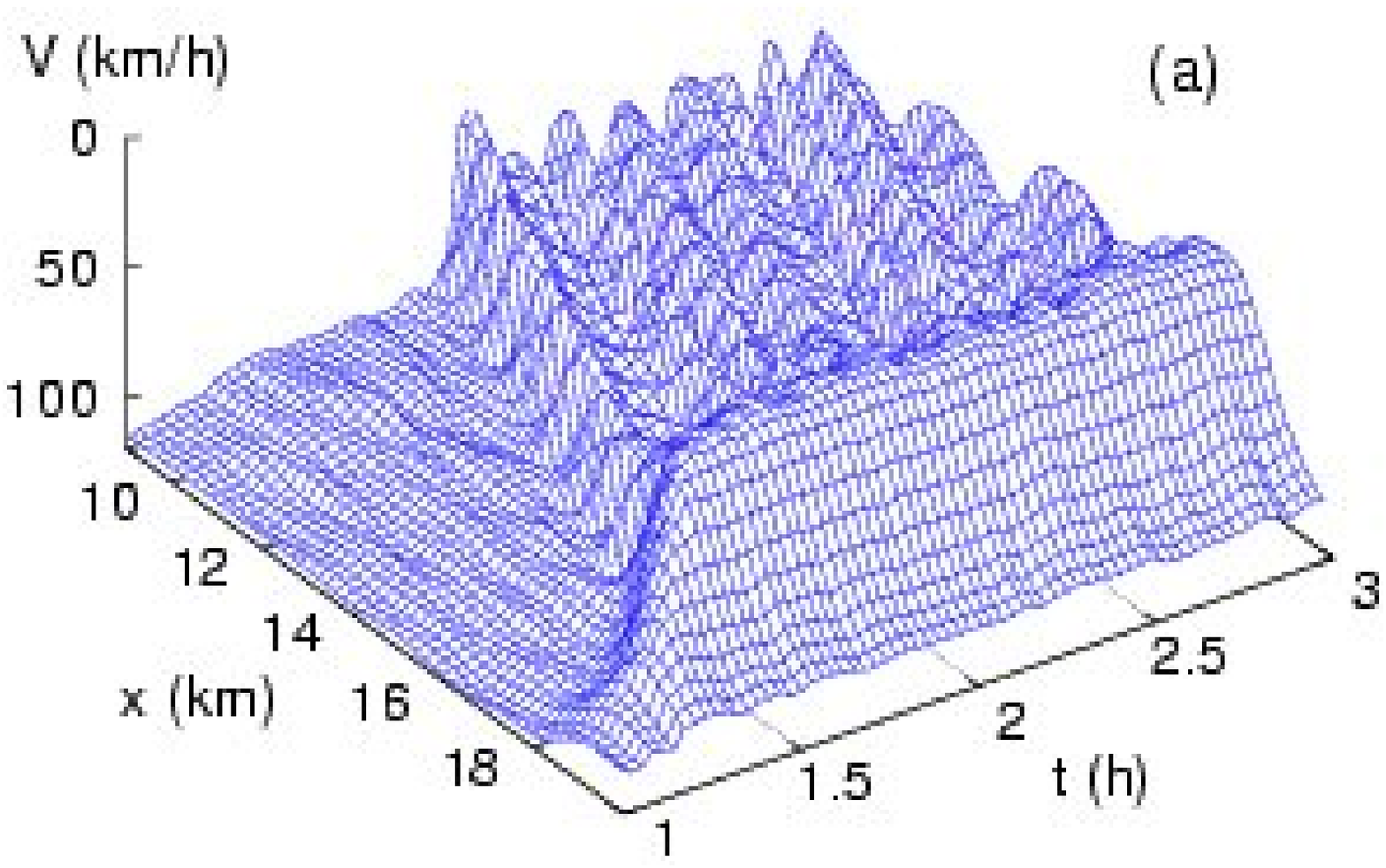}
 \includegraphics[width=65mm]{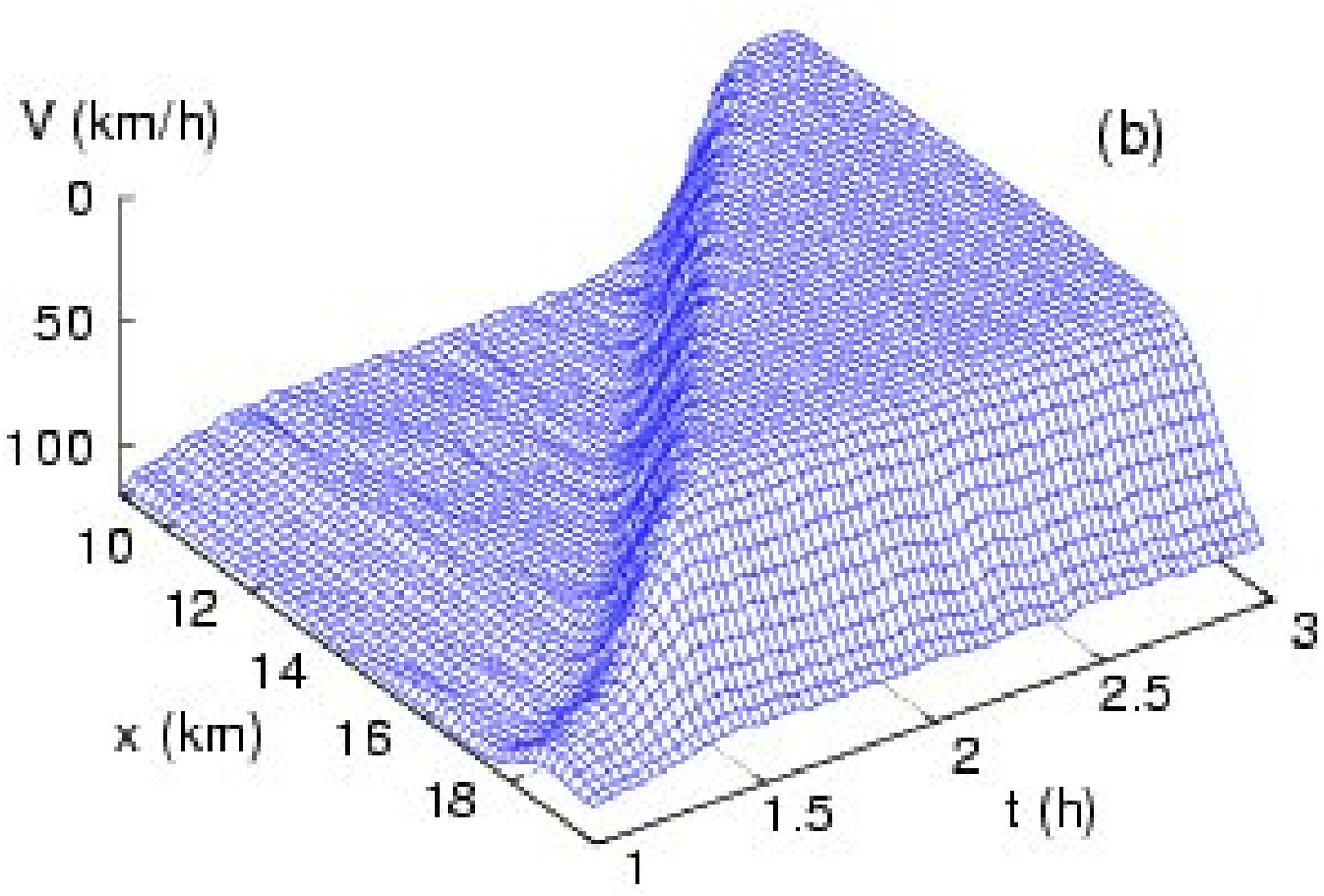} 
 \includegraphics[width=65mm]{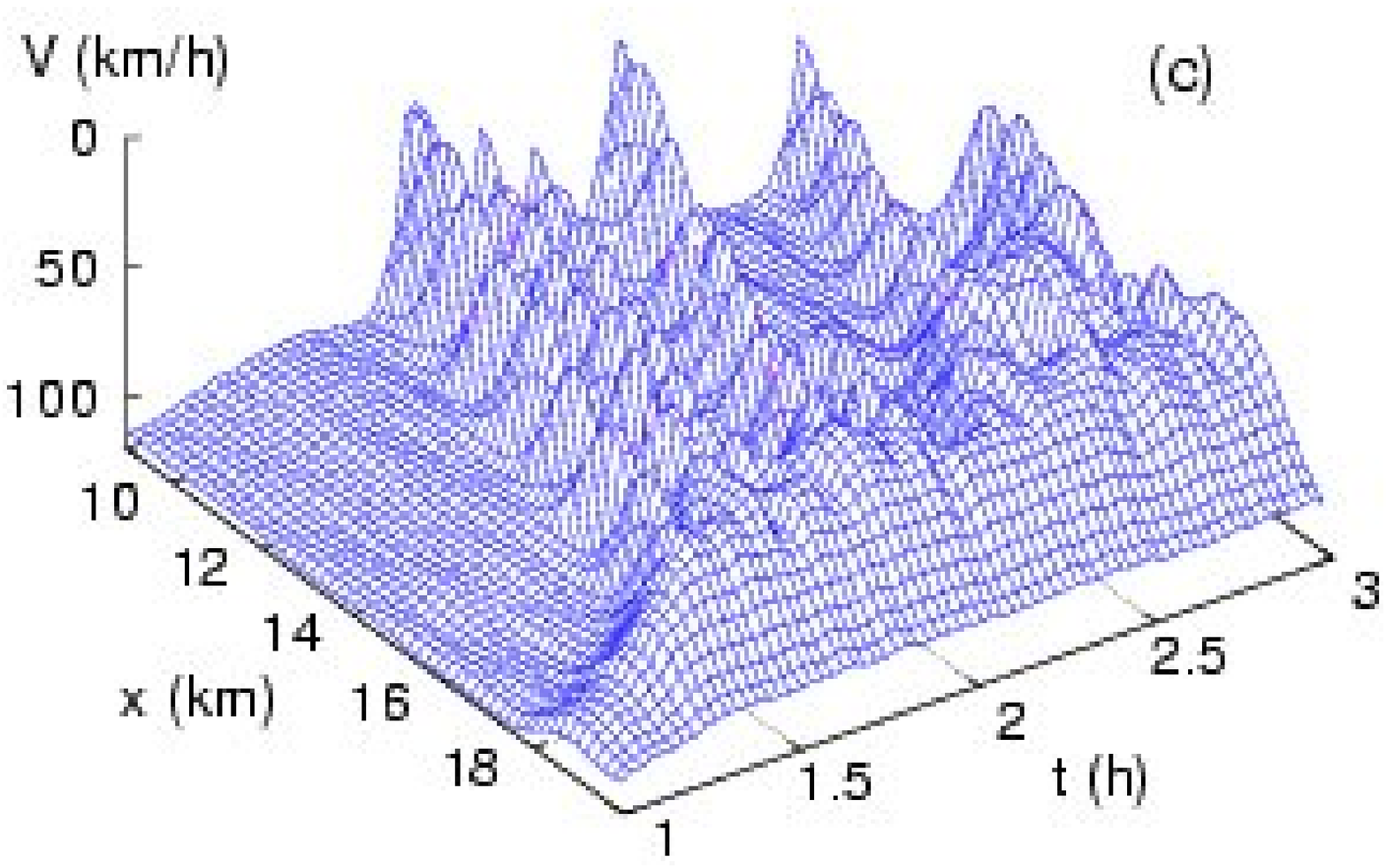}
 \includegraphics[width=65mm]{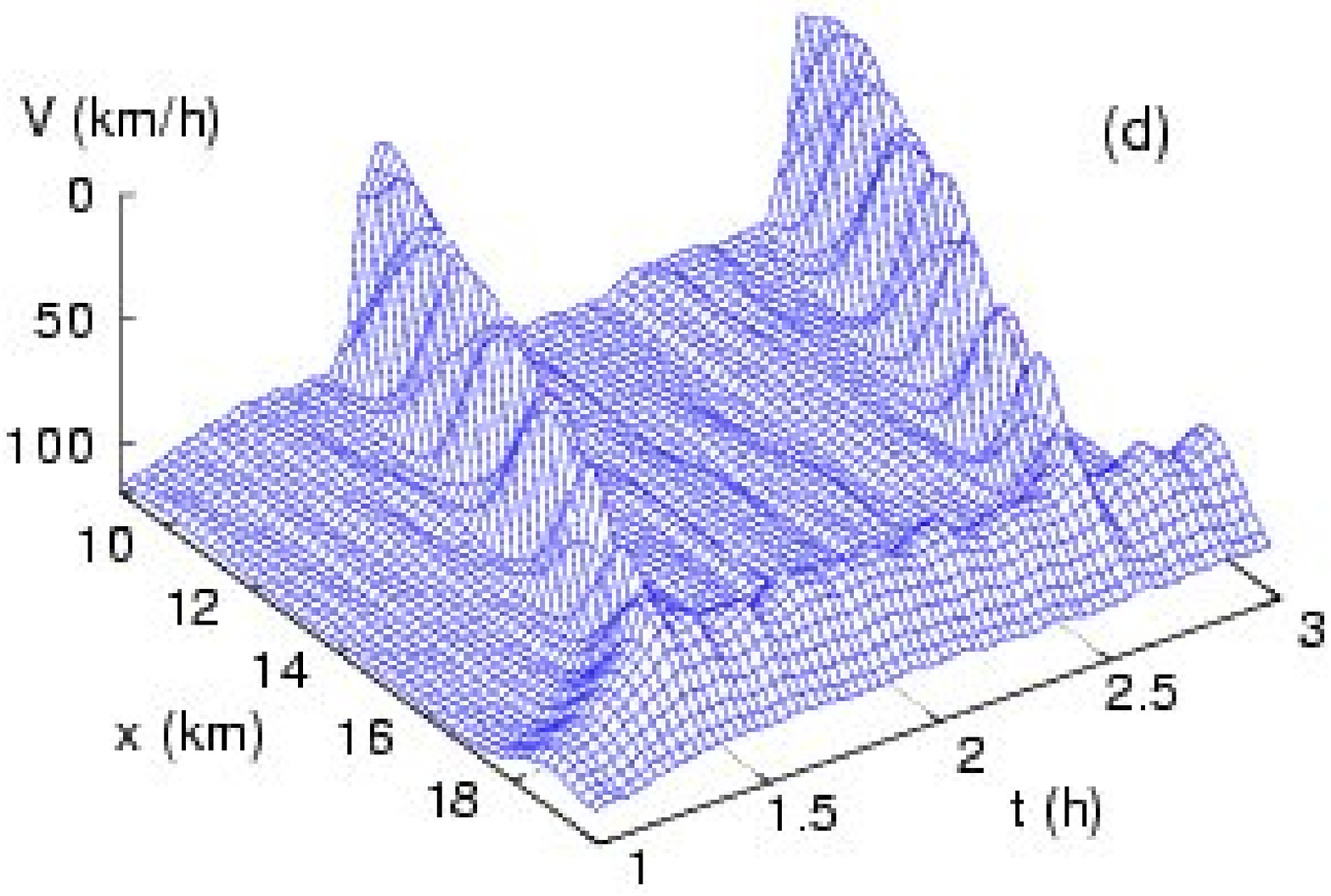}
  \end{center}
\caption{\label{figTrafficStates}Spatiotemporal dynamics of typical 
  traffic states of the phase diagram of Fig.
  \ref{figPhaseDiagramTrafficStates}.  (a) The special case of the IDM
($n_a=1$,
  $T'=0\:\mathrm{s}$) leads to oscillatory congested traffic (OCT);
  (b) $n_a=5$ anticipated vehicles and a reaction time 
$T'=0.9\:\mathrm{s}$ leads to homogeneous congestion (HCT); even larger
reaction times lead to (c) triggered stop-and-go traffic (TSG) ($n_a=5$,
  $T'=1.1\:\mathrm{s}$) or (d) a combination of moving (MLC) and pinned
localized clusters (PLC) ($n_a=6$, $T'=1.2\:\mathrm{s}$).
All data are smoothed with half-widths of 1 min and 0.4 km, respectively.
  }
\end{figure}

We have simulated the open system with $n_a=1$ to $n_a=7$. For each
value of $n_a$, we have varied the reaction time in steps of
0.05~s. Figure \ref{figTrafficStates} shows typical examples of the
spatiotemporal patterns occuring in the simulations. By associating
qualitatively different simulation results with different dynamical
phases, we obtained a phase diagram in the space spanned by $n_a$ and
$T'$ (see Fig. \ref{figPhaseDiagramTrafficStates}).  The different
states were determined using smoothed velocity data, as shown in
Fig.~\ref{figTrafficStates}.

Specifically, a congested state may be either localized (localized cluster,
LC) or extended (extended congested traffic, ECT). The criterion to
discriminate between these two types of congested traffic is the width
of the congested region which, for LCs, is constant (and typically
less than 1 km), while the width of ECT is variable and depend in
particular on the inflow. The transition between LC and ECT is
slightly hysteretic. Furthermore, there are transitions from both LC
and ECT to free traffic which are hysteretic as well.

\begin{figure}
  \begin{center}
  \includegraphics[width=100mm]{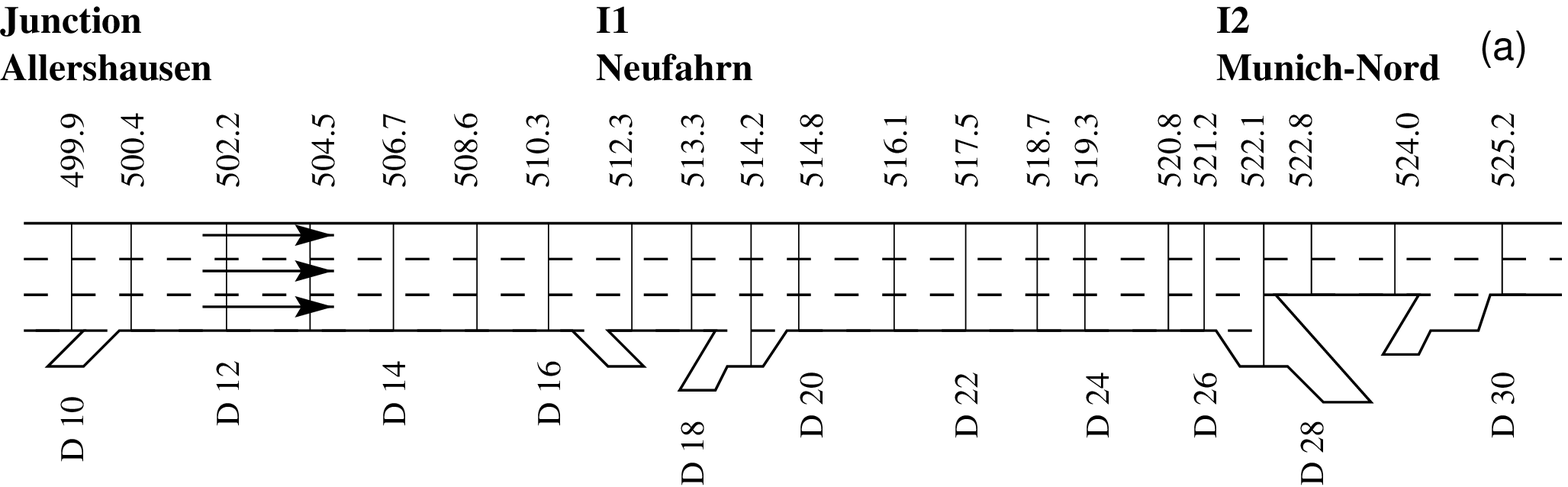}
  \includegraphics[width=65mm]{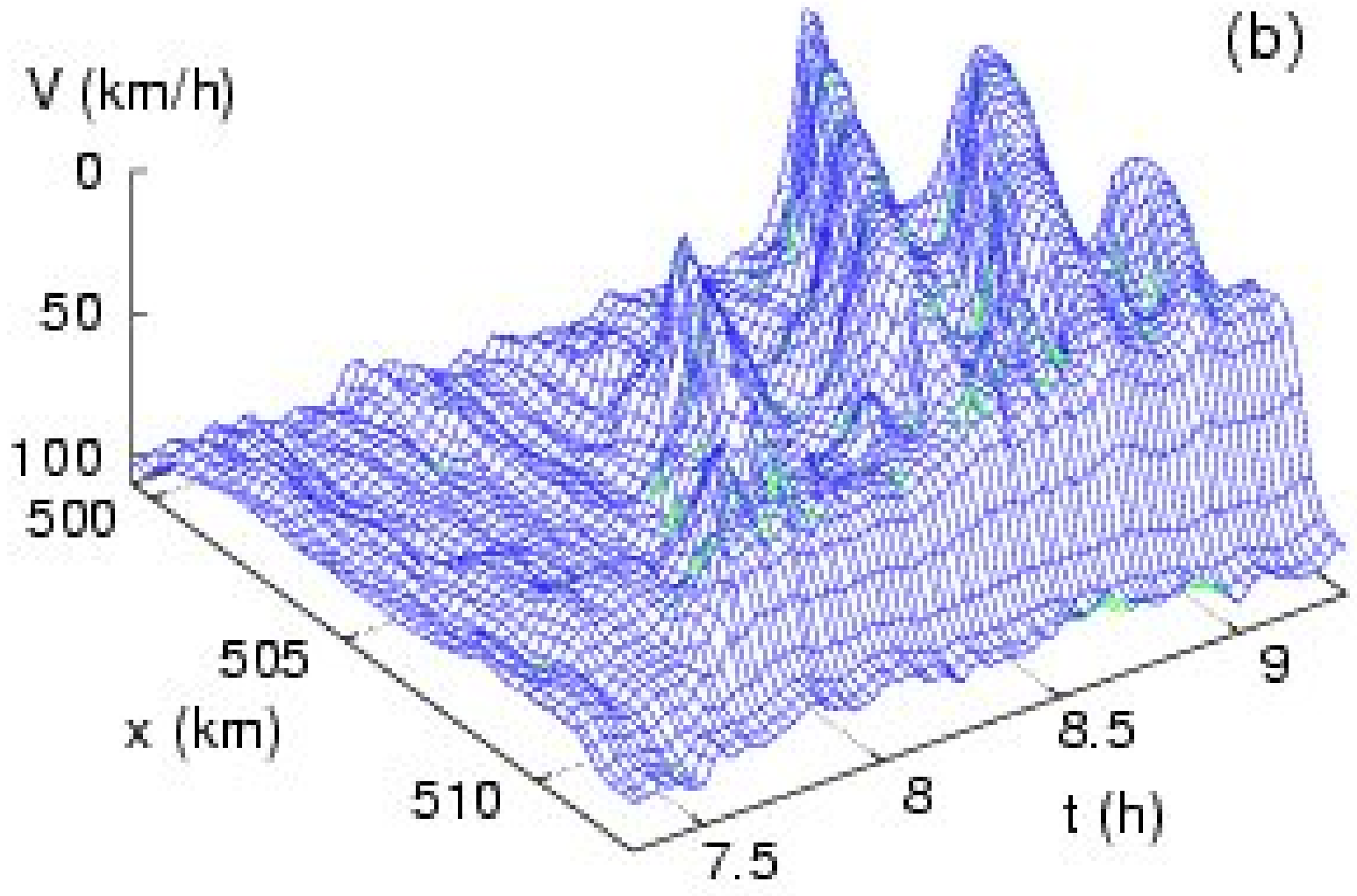}
  \includegraphics[width=65mm]{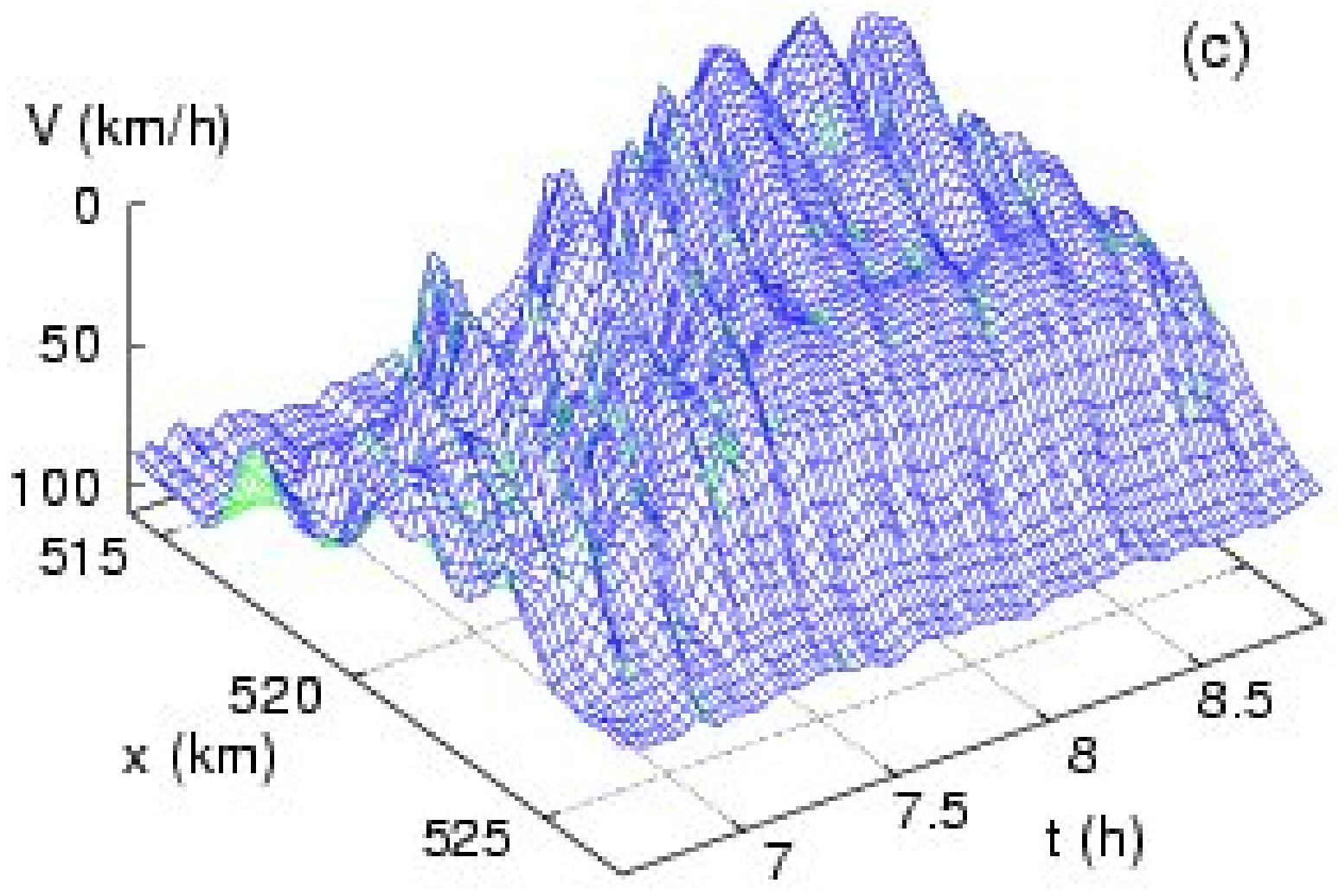}
  \end{center}
\caption{\label{figEmpStates}Spatiotemporal dynamics of  
  congested traffic on the German freeway A9 South near Munich.  (a)
  Sketch of the freeway. (b) Stop-and-go traffic (TSG) caused by the
  intersection I1 (cf. Fig. \ref{figTrafficStates}c). (c) Oscillatory
  congested traffic (OCT) caused by the intersection I2
  (cf. Fig. \ref{figTrafficStates}a). The spatiotemporal data were
  obtained from one minute data at the detector positions indicated in
  (a), using the adaptive smoothing method \cite{Treiber-smooth} with
  smoothing half-widths of 1 min and 0.4 km, respectively. }
\end{figure}

Within ECT, there exist three dynamical phases separated by continuous
phase transitions. As order parameter to distinguish between
homogeneous congested traffic (HCT,
cf. Fig. \ref{figTrafficStates}(b)) and oscillating congested traffic
(OCT, cf. Fig. \ref{figTrafficStates}(a) and \ref{figEmpStates}(c)) we
have used the variance $\theta$ of the temporal velocity variations in the
congested region sufficiently upstream of the bottleneck, where it
is essentially constant with respect to space and time. While, in the
case of HCT, $\theta$ depends mainly on the fluctuating forces and remains
below 1 (m/s)$^2$, it jumps to more than 100 (m/s)$^2$ and essentially
becomes independent of the fluctuation strength in the case of OCT.  The
third dynamical ECT phase are triggered stop-and go waves (TSG,
cf. Fig. \ref{figTrafficStates}(c) and \ref{figEmpStates}(b)).  In
contrast to OCT, TSG states reach the free branch of the fundamental
diagram, i.e. there are uncongested areas between the congested
ones. Nevertheless, the OCT and TSG states are hard to distinguish as
they are not separated by a hysteretic phase transition.

Within localized clusters, we have observed a sharp transition between
LCs moving upstream at a constant velocity of about $v_g=-15$ km/h
(MLC), and clusters fixed at the bottleneck (PLC). In contrast to the
IDM, we observed a coexistence of both localized dynamical phases
(Fig. \ref{figTrafficStates}(d)), as required by observations
\cite{empiricalPhaseDia-martin2}.

\begin{figure}
  \begin{center}
    \includegraphics[width=65mm]{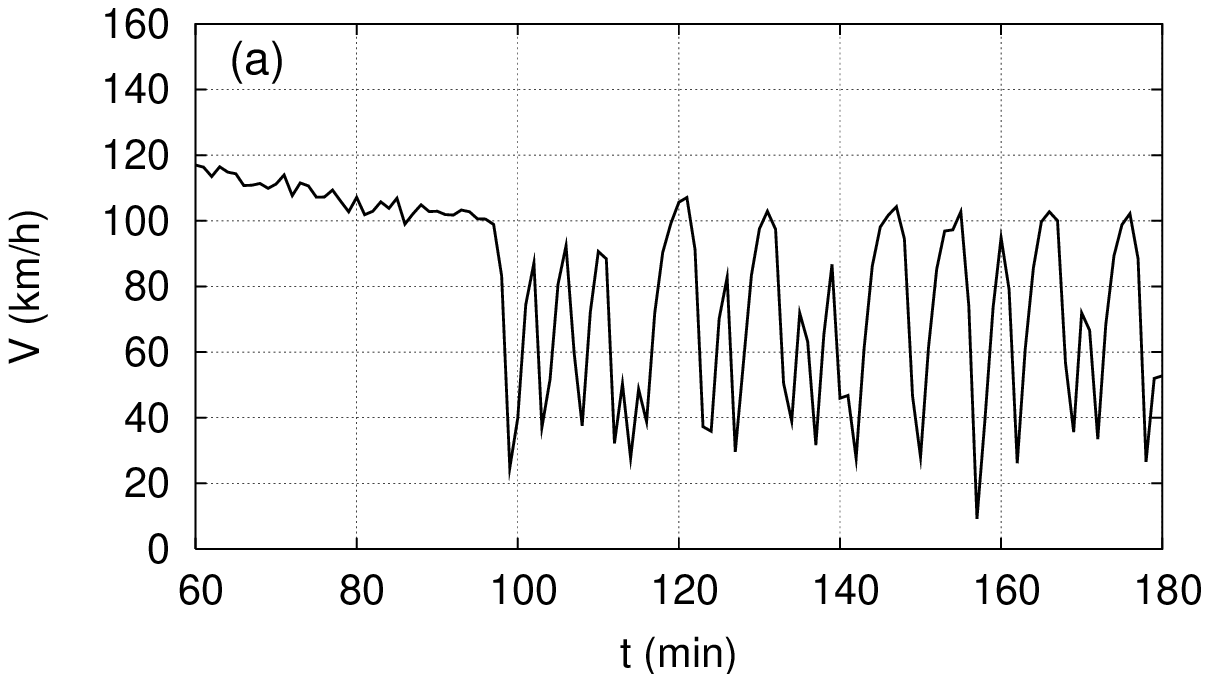}
    \includegraphics[width=65mm] {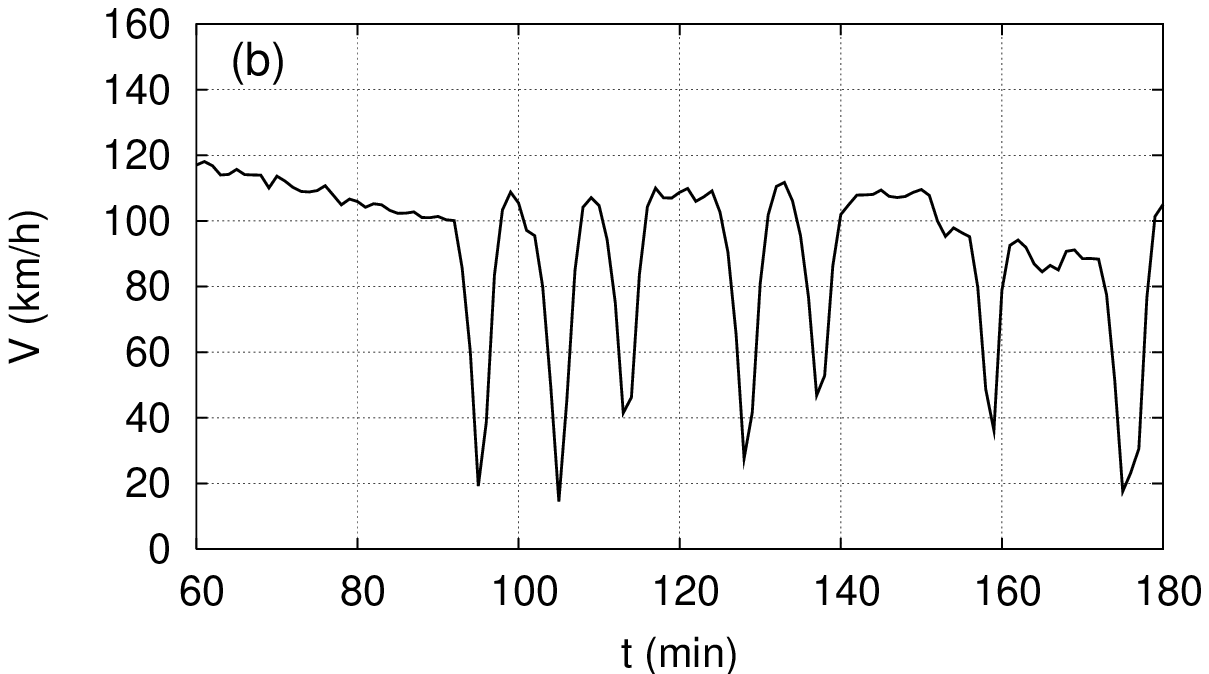}
  \end{center}
\caption{\label{figVirtualDetectorVel1}Velocity time series 
  by a virtual detector at $x=14$ km for (a) $n_a=1$,
  $T'=0\:\mathrm{s}$, and (b) $n_a=5$, $T'=1.1\:\mathrm{s}$. Notice
  the increase of the oscillation wavelength in the scenario (b) with
  anticipation.}
\end{figure}

Let us point out that our system is markedly different from the open
system proposed by Kolomeisky et. al \cite{Kolomeisky-98} where all
phases are triggered by boundaries rather than by a bottleneck and
which essentially contains only free traffic (FT) and the HCT state.
Specifically, one can associate the `high-density' state of
\cite{Kolomeisky-98} with HCT, while the `maximal-current' state
corresponds to the outflow region from HCT and finally the
`low-density' state corresponds to FT.

All of the above dynamic phases can be either reduced by varying the
inflow and the bottleneck capacity \cite{Phase,GKT,Opus}, or by
varying the model parameters, which we do in the following.  The left
lower corner of Fig.~\ref{figPhaseDiagramTrafficStates} corresponds to
the special case of the IDM, i.e. to the case of zero reaction time
($T'=0$) and consideration of the immediate front vehicle only
($n_a=1$).  In this case, the simulation results in OCT for the
boundary conditions specified before, see Figures
\ref{figTrafficStates}(a),  and \ref{figVirtualDetectorVel1}(a).
  
Varying $n_a$ and $T'$ leads to the following main results: 
\begin{enumerate}
\item[(i)] Traffic stability increases drastically, when the spatial anticipation
is increased from $n_a=1$ to $6$, while the stability remains
essentially unchanged for $n_a>6$. 
\item[(ii)] For a sufficiently large number of anticipated
vehicles, the congestion pattern becomes stable corresponding to homogeneous congested
traffic (HCT) as shown in Fig.~\ref{figTrafficStates}(b). Thus, different
traffic states can be produced not only by varying the bottleneck
strength as in the phase diagram proposed in Ref. \cite{Phase}, but
also by varying model parameters that influence stability.
\item[(iii)] Increasing the reaction time $T'$ destabilizes traffic and
finally leads to crashes. 
\item[(iv)] The other dynamic congested traffic states of the phase diagram
presented in Ref. \cite{Phase} are found as well, specifically,
triggered stop-and-go waves (TSG), cf. Figures
\ref{figTrafficStates}(c) and an interesting combination of moving
localized clusters (MLC) and pinned localized clusters (PLC), see
Fig.~\ref{figTrafficStates}(d). 
\item[(v)] The results are robust against variations of the stochastic HDM
parameters or when the correlated noise is replaced by white
acceleration noise.
\end{enumerate}

We have compared the spatiotemporal dynamics with traffic data from
the German freeway A9 South near Munich.  For this freeway, aggregated
detector data (vehicle counts and average velocities for one-minute
intervals) are available for each lane at the locations indicated in
Fig. \ref{figEmpStates} (a).
The plots (b) and (c) of Figure \ref{figEmpStates} show the
spatiotemporal dynamics of the local velocity for triggered
stop-and-go traffic (TSG) and oscillating congested traffic (OCT),
respectively. In both cases, these states were caused by intersections
acting as bottlenecks.

To obtain the spatiotemporal dynamics shown in these plots, we have
applied the adaptive smoothing method (ASM) \cite{Treiber-smooth} to
the lane-averages of the velocity data.  The smoothing times and
lengths of the ASM are set to the values used for the averaging filter
in the simulation, i.e., to 1 min and 0.4 km, respectively.  The other
ASM parameters were set to $v_{c1}=30$ km/h, $v_{c2}=60$ km/h,
$c\sub{free}=80$ km/h, and $c\sub{cong}=-15$ km/h
\cite{Treiber-smooth}.  We have checked that the result was essentially
unchanged when changing any of the first three parameters by factors
between 0.7 and 1.5. In contrast, changes of the last parameter
$c\sub{cong}$ re\-presenting the propagation velocity of collective
structures in congested traffic influence the result.  Artificial
shifts of the congested structures are observed if $c\sub{cong}$ is
outside a range of about $[-17 $ km/h, $-14$ km/h]. This means,
besides making spatiotemporal plots, the ASM can be used to determine
the propagation velocity $c\sub{cong}$.

By comparing the simulation results \ref{figTrafficStates}(a) and (c)
with the traffic data shown in \ref{figEmpStates}(c) and (b),
respectively, one sees a qualitative agreement of the spatiotemporal
dynamics in many respects.  Particularly, 
\begin{enumerate}
\item[(i)] the congestion pattern
is triggered by a bottleneck, 
\item[(ii)] the downstream front of the
congestion pattern is stationary and located at the position of the
bottleneck, 
\item[(iii)] traffic is essentially non-oscillatory in a region
of about 1 km width near the bottleneck (this is sometimes called the
`pinch region' \cite{KernerPinch}), 
\item[(iv)] further upstream, the
congested traffic consists of stop-and-go waves propagating upstream
at a constant velocity $c\sub{cong}$, 
\item[(v)] the period $\tau$ of the
oscillations is variable.
\end{enumerate}

Isolated and coexisting MLCs and PLCs as in
Fig. \ref{figTrafficStates}(d) have been observed in traffic data as
well \cite{empiricalPhaseDia-martin2}. Particularly, PLC and MLC
states are shown in Fig.~6 and Fig.~10 of
Ref. \cite{empiricalPhaseDia-martin2}.  HCT states and even combined
states of one or more PLCs with one or more isolated MLCs were also
observed in \cite{empiricalPhaseDia-martin2}.

In addition to these qualitative aspects, there exists a nearly
quantitative agreement with respect to (i) the propagation velocity
$c\sub{cong}=-15$ km/h, and (ii) the range of the oscillation periods
$\tau$ between 6 min and 40 min (notice that many car-following models
yield too short periods).  The latter point is illustrated by
comparing Fig.  \ref{figTrafficStates}(a) (no reaction time and no
anticipation) and Fig. \ref{figTrafficStates}(c) (finite reaction time
and anticipation) with the data, Fig.~\ref{figEmpStates}.

In summary, we have shown that the destabilizing effects of finite
reaction times can be compensated to a large extent by spatial and
temporal anticipation such that the resulting stability and dynamics
are similar to the case of the IDM with zero anticipation and reaction
time. However, besides stability issues, the HDM simulation results
agree better with empirical traffic data in the following aspects: (i)
Compared to the underlying IDM, the HDM simulation shows larger
oscillation periods in the case of OCT and, generally, lower velocity
gradients. (ii) Coexisting PLCs and MLCs are observed both in the HDM
and in real traffic data \cite{empiricalPhaseDia-martin2}, but not in
the IDM.  (iii) Near the bottleneck, the HDM regularly produces
traffic of relatively high flow and density ('pinch region').

\section{\label{sec:diss}Discussion}

Finite reaction times and errors in estimating the input variables are
clearly essential factors of driver behaviour affecting the
performance and stability of vehicular traffic. However, these aspects
are rarely considered in physics-oriented traffic
modelling. Nevertheless, the simple models used by physicists such as
the optimal-velocity model and its generalizations
\cite{Bando,Davis-02}, the velocity-difference model
\cite{Jiang-vdiff}, or the IDM allow to describe many, particularly
macroscopic, aspects of traffic dynamics such as the spatiotemporal
dynamics of the various types of traffic congestion, the propagation
of stop-and-go traffic, or even the scattering of flow-density data
points of `synchronized traffic' \cite{Helb-crit}.

The question arises why, despite their obvious shortcomings, these
models work so well. This question became more pressing after
it turned out that all of the above models (including the IDM) produce
unrealistic dynamics and crashes when simulating these models
with realistic reaction times (of the order of 1 s).

In this work, we have shown that the destabilizing effects of reaction
times and estimation errors can be compensated for by spatial and
temporal anticipations: One obtains essentially the same longitudinal
dynamics, which explains the good performance of the underlying simple
models.  In order to put this balance of stabilizing and destabilizing
effects into a more general context, we have formulated the
human-driver model (HDM) as a meta-model that can be used to extend a
wide class of car-following models, where the acceleration depends
only on the positions, velocities and accelerations of the own and the
preceding vehicle.  By applying the HDM extensions to the IDM, we have
provided quantitative details of the balance conditions and the
remaining differences in the dynamics. This involves validity criteria
for the applicability of simpler physics-oriented car-following
models.

Since the basic models have both the advantages and limitations of
adaptive cruise control (ACC) systems, one can investigate the impact
of ACC vehicles on the capacity and stability of the overall traffic
simply by simulating a mixture of, e.g., IDM and HDM
vehicles. Furthermore, one gets the nontrivial result that
hypothetical future traffic consisting predominantly of automated
vehicles will exhibit macroscopic dynamics similar to that of the
actual traffic, although the driving strategy would be markedly
different.

While finite reaction times have been investigated for more than 40
years \cite{Helb-Opus} the HDM-IDM combination is, to our knowledge,
the first car-following model allowing accident-free driving at
realistic accelerations in all traffic situations for reaction times
of the order of and even exceeding the time headway.  A closer look at
quantitative features of stop-and-go traffic or oscillations shows
that, compared to simulations of the original IDM, the HDM extensions
reduce the gradients of transitions between free and congested traffic
and increase the wavelengths of stop-and-go waves, in agreement with
empirical data. This suggests that multi-anticipation is an essential
aspect of the driver behaviour.

A comparison of the stochastic HDM expressions for imperfect
estimation capabilities with other stochastic micromodels is in
order. While fluctuating terms were first introduced to traffic models
more than 20 years ago \cite{Gipps81}, the most prominent example of
stochastic traffic models are cellular automata (CA) of the
Nagel-Schreckenberg type \cite{Nagel-S} and extensions thereof.  There
is, however, a qualitative difference compared with most continuous
models: Fluctuation terms change the qualitative dynamics of many CA
models. Therefore, they must be carefully chosen to yield plausible
results. In contrast, the qualitative dynamics typically remains the
same when fluctuations are added to car-following models {\it via} the
HDM extensions.  Having modelled the estimation errors by a stochastic
Wiener process with a finite correlation time, we have included the
persistence of estimation errors for a certain time interval.

The phase diagram shown in Fig.~\ref{figPhaseDiagramTrafficStates}
contained qualitatively the same spatiotemporal congested states as
found in Refs. \cite{Phase,Opus}. At first sight, this seems
surprising. Besides using a different model, the control parameters
making up the phase space were extrinsic in the previous work, while
the phase space is spanned by intrinsic model parameters in the
present work. A closer look at the analytic expressions for the phase
boundaries \cite{Phase} containing both extrinsic flow parameters and
intrinsic stability limits indicates that variations of both kinds of
control parameters can lead to phase transitions.  

From a control-theoretical point of view, the HDM extensions
 implement a continuous
response to delayed and noisy input stimuli. Alternatively, human
driving behaviour can be modelled by so-called action-point models,
where the response changes discontinuously whenever certain boundaries
in the space spanned by the input stimuli are crossed
\cite{Wiedemann,Wagner-03,Eissfeldt03}, but these thresholds
cannot easily confirmed by empirical data.  
 
Finally, it should be mentioned that, in this work, we have considered
only longitudinal aspects of human driving (acceleration and
deceleration) and implemented only identical driver-vehicle units.
Platooning effects due to different driving styles and the remarkable
ability of human drivers to safely and smoothly change lanes even in
congested conditions are the topic of a forthcoming paper.

\textbf{Acknowledgments:}
The authors would like to thank for partial support by the
DFG project He 2789/2-2 and the Volkswagen AG within the BMBF project
INVENT.




\begin{thebibliography}{10}
\expandafter\ifx\csname url\endcsname\relax
  \def\url#1{\texttt{#1}}\fi
\expandafter\ifx\csname urlprefix\endcsname\relax\def\urlprefix{URL }\fi

\bibitem{Helb-Opus}
D.~Helbing, Traffic and related self-driven many-particle systems, Review of
  Modern Physics 73 (2001) 1067--1141.

\bibitem{Brackstone-hist}
M.~Brackstone, M.~McDonald, Car-following: a historical review, Transp. Res. F
  2 (1999) 181--196.

\bibitem{Holland97}
E.~Holland, A generalised stability criterion for motorway traffic, Transp.
  Res. B 32 (1998) 141--154.

\bibitem{Helb-crit}
D.~Helbing, M.~Treiber, Critical discussion of ''synchronized flow'',
  Cooper@tive Tr@nsport@tion Dyn@mics 1 (2002) 2.1--2.24, (Internet Journal,
  {\tt www.TrafficForum.org/journal}).

\bibitem{nagel-wagner-vdiff}
K.~Nagel, P.~Wagner, R.~Woesler, Still flowing: old and new approaches for
  traffic flow modeling, Operations Research 51 (2003) 681--710.

\bibitem{Kerner-traffic}
B.~S. Kerner, The Physics of Traffic. Empirical Freeway Pattern Features,
  Engineering Applications, and Theory, Understanding Complex Systems,
  Springer, 2004.

\bibitem{chowdhury-report}
D.~Chowdhury, D.~Santen, A.~Schadschneider, Statistical physics of vehicular
  traffic and some related systems, Physics Reports 329 (2000) 199--329.

\bibitem{nagatani-report}
T.~Nagatani, The physics of traffic jams, Reports of Progress in Physics 65
  (2002) 1331--1386.

\bibitem{Treiber-aut}
M.~Treiber, D.~Helbing, Microsimulations of freeway traffic including control
  measures, Automatisierungstechnik 49 (2001) 478--484.

\bibitem{marsden-ACC}
G.~Marsden, M.~McDonald, M.~Brackstone, Towards an understanding of adaptive
  cruise control, Transportation Research C 9 (2001) 33--51.

\bibitem{Newell}
G.~Newell, Nonlinear effects in the dynamics of car following, Operations
  Research 9 (1961) 209.

\bibitem{Bando-98}
M.~Bando, K.~Hasebe, K.~Nakanishi, A.~Nakayama, Analysis of optimal velocity
  model with explicit delay, Phys. Rev. E 58 (1998) 5429.

\bibitem{Davis-02}
L.~Davis, Modifications of the optimal velocity traffic model to include delay
  due to driver reaction time, Physica A 319 (2002) 557.

\bibitem{Bando}
M.~Bando, K.~Hasebe, A.~Nakayama, A.~Shibata, Y.~Sugiyama, Dynamical model of
  traffic congestion and numerical simulation, Phys. Rev. E 51 (1995)
  1035--1042.

\bibitem{BandoDelay-02}
L.~Davis, Comment on 'analysis of optimal velocity model with explicit delay',
  Phys. Rev. E 66 (2002) 038101.

\bibitem{Kno01}
W.~Knospe, L.~Santen, A.~Schadschneider, M.~Schreckenberg, Human behaviour as
  origin of traffic phases, Phys. Rev. E 65 (2001) 015101.

\bibitem{Tilch-TGF99}
B.~Tilch, D.~Helbing, Evaluation of single vehicle data in dependence of the
  vehicle-type, lane, and site, in: D.~Helbing, H.~Herrmann, M.~Schreckenberg,
  D.~Wolf (Eds.), Traffic and Granular Flow '99, Springer, Berlin, 2000, pp.
  333--338.

\bibitem{Kno02-data}
W.~Knospe, L.~Santen, A.~Schadschneider, M.~Schreckenberg, Single-vehicle data
  of highway traffic: Microscopic description of traffic phases, Phys. Rev. E
  65 (2002) 056133.

\bibitem{green-reactionTimes}
M.~Green, "{H}ow {L}ong {D}oes {I}t {T}ake to {S}top?" {M}ethodological
  analysis of driver perception-brake times, Transportation Human Factors 2
  (2000) 195--216.

\bibitem{Lenz-Wagner}
H.~Lenz, C.~Wagner, R.~Sollacher, Multi-anticipative car-following model,
  European Physical Journal B7 (1998) 331--335.

\bibitem{Eissfeldt03}
N.~Eissfeldt, P.~Wagner, Effects of anticipatory driving in a traffic flow
  model, European Physical Journal B 33 (2003) 121--129.

\bibitem{CA_limitedAcc}
H.~K. Lee, R.~Barlovic, M.~Schreckenberg, D.~Kim, Mechanical restriction versus
  human overreaction triggering congested traffic states, Physical Review
  Letters 92 (2004) 238702.

\bibitem{Gipps81}
P.~G. Gipps, A behavioural car-following model for computer simulation, Transp.
  Res. B 15 (1981) 105--111.

\bibitem{Wagner-03}
P.~Wagner, I.~Lubashevsky, Empirical basis for car-following theory
  development, cond-mat/0311192.

\bibitem{Jiang-vdiff}
R.~Jiang, Q.~Wu, Z.~Zhu, Full velocity difference model for a car-following
  theory, Phys. Rev. E 64 (2001) 017101.

\bibitem{Opus}
M.~Treiber, A.~Hennecke, D.~Helbing, Congested traffic states in empirical
  observations and microscopic simulations, Physical Review E 62 (2000)
  1805--1824.

\bibitem{Lub03}
I.~Lubashevsky, P.~Wagner, R.~Mahnke, Bounded rational driver models, European
  Physical Journal B 32 (2003) 243.

\bibitem{Mahnke03}
I.~Lubashevsky, P.~Wagner, R.~Mahnke, Rational-driver approximation in
  car-following theory, Phys. Rev. E 68 (2003) 056109.

\bibitem{Brockfeld-Benchmark}
E.~Brockfeld, R.~K\"uhne, P.~Wagner, Towards benchmarking microscopic traffic
  flow models, in: W.~M\"ohlenbrink, M.~Bargende, U.~Hangleiter, U.~Martin
  (Eds.), Networks for Mobility 2002, FOVUS: International Symposium, September
  18-20, Stuttgart, Stuttgart, 2002, pp. 321--331.

\bibitem{Treiber-TGF99}
M.~Treiber, A.~Hennecke, D.~Helbing, Microscopic simulation of congested
  traffic, in: D.~Helbing, H.~Herrmann, M.~Schreckenberg, D.~Wolf (Eds.),
  Traffic and Granular Flow '99, Springer, Berlin, 2000, pp. 365--376.

\bibitem{remark-acc}
Inclusion of the acceleration $a_{\alpha-1}(t)$ of the preceding vehicle to the
  input variables is straightforward allowing for micromodels that model
  reactions to braking lights.

\bibitem{IDMM}
M.~Treiber, D.~Helbing, Memory effects in microscopic traffic models and wide
  scattering in flow-density data, Phys. Rev. E 68 (2003) 046119.

\bibitem{remark-bias}
Generally, the estimation error concludes a systematic bias as well. We found
  that our model is very robust with respect to reasonable biases in distance
  and velocity-difference estimates.

\bibitem{Nagel-S}
K.~Nagel, M.~Schreckenberg, A cellular automaton modell for freeway traffic, J.
  Phys. I France 2 (1992) 2221--2229.

\bibitem{Gardiner}
C.~Gardiner, Handbook of Stochastic Methods, Springer, N.Y., 1990.

\bibitem{Knospe}
W.~Knospe, L.~Santen, A.~Schandschneider, M.~Schreckenberg, Disorder effects in
  cellular automata for two-lane traffic, Physica A 265 (1998) 614--633.

\bibitem{MOBIL-rostock}
M.~Treiber, D.~Helbing, {R}ealistische {M}ikrosimulation von {S}trassenverkehr
  mit einem einfachen {M}odell, in: D.~Tavangarian, R.~Gr\"utzner (Eds.), ASIM
  2002, Tagungsband 16. Symposium Simulationstechnik, Rostock, 2002, pp.
  514--520.

\bibitem{Treiber-smooth}
M.~Treiber, D.~Helbing, Reconstructing the spatio-temporal traffic dynamics
  from stationary detector data, Cooper@tive Tr@nsport@tion Dyn@mics 1 (2002)
  3.1--3.24, (Internet Journal, {\tt www.TrafficForum.org/journal}).

\bibitem{empiricalPhaseDia-martin2}
M.~Sch\"onhof, D.~Helbing, Empirical {F}eatures of {C}ongested {T}raffic
  {S}tates and their {I}mplications for {T}raffic {M}odeling, cond-mat/0408138.

\bibitem{Kolomeisky-98}
A.~B. Kolomeisky, G.~M. Sch{\" u}tz, E.~B. K. J.~P. Straley, Phase diagram of
  one-dimensional driven lattice gases with open boundaries, Journal of Physics
  A Mathematical General 31 (1998) 6911--6919.

\bibitem{Phase}
D.~Helbing, A.~Hennecke, M.~Treiber, Phase diagram of traffic states in the
  presence of inhomogeneities, Phys. Rev. Lett. 82 (1999) 4360--4363.

\bibitem{GKT}
M.~Treiber, A.~Hennecke, D.~Helbing, Derivation, properties, and simulation of
  a gas-kinetic-based, non-local traffic model, Phys. Rev. E 59 (1999)
  239--253.

\bibitem{KernerPinch}
B.~Kerner, Experimental features of self-organization in traffic flow, Phys.
  Rev. Lett. 81 (1998) 3797--3800.

\bibitem{Wiedemann}
R. Wiedemann, {\it Simulation des Stra{\ss}enverkehrsflusses}, Schriftenreihe
  des IfV Vol. {\bf 8}, Institut f\"ur Verkehrswesen, Universit{\"at} Karlsruhe
  (1974).

\end{thebibliography}


\end{document}